\documentclass{PoS}

\title{Swift/XRT observations of newly discovered INTEGRAL sources}

\ShortTitle{Swift/XRT observations of newly discovered INTEGRAL sources}

\author{\speaker{R. Landi}$^{a}$, L. Bassani$^{a}$, A. Bazzano$^{b}$, M. Fiocchi$^{b}$
and A. J. Bird$^{c}$\\
\llap{$^{a}$} INAF -- IASF Bologna\\
\llap{$^{b}$} INAF -- IAPS Rome\\
\llap{$^{c}$} University of Southampton\\
E-mail: \email{landi@iasfbo.inaf.it}, \email{bassani@iasfbo.inaf.it},
\email{angela.bazzano@iaps.inaf.it}, \email{mariateresa.fiocchi@iasf-roma.inaf.it},
\email{A.J.Bird@soton.ac.uk}}

\abstract{With respect to the recent \emph{INTEGRAL}/IBIS 9--year Galactic Hard X--ray Survey
(Krivonos et al. 2012), we use archival \emph{Swift}/XRT observations in conjunction with
multi--wavelength information to discuss the counterparts of a sample of newly discovered objects.
The X--ray telescope (XRT, 0.3--10 keV) on board \emph{Swift}, thanks to its few arcseconds source
location accuracy, has been proven to be a powerful tool with which the X--ray counterparts to these
IBIS sources can be searched for and studied. In this work, we present the outcome of this analysis
by discussing four objects (SWIFT J0958.0--4208, SWIFT J1508.6--4953, IGR J17157--5449, and 
IGR J22534+6243) having either X--ray data of sufficient quality to perform a reliable
spectral analysis or having interesting multiwaveband properties.
We find that SWIFT J1508.6--4953 is most likely a Blazar, while IGR J22534+6243 is probably a HMXB.
The remaining two objects may be contaminated by nearby X--ray sources and their class can be 
inferred only by means of optical follow--up observations of all likely counterparts.}

\FullConference{An INTEGRAL view of the high-energy sky (the first 10 years)"
9th INTEGRAL Workshop and celebration of the 10th anniversary of the launch,\\
		October 15-19, 2012\\
		Bibliotheque Nationale de France, Paris, France}

\begin{document}

\section{Introduction}
X--ray follow-up observations are an important part of the \emph{INTEGRAL} survey 
process as they enable us to pinpoint the optical counterpart of unidentified objects,
proceed with their classification and study their X--ray properties.
Here we describe the results of follow--up observations of four still 
unclassified sources reported in the \emph{INTEGRAL}/IBIS 9--year Galactic Hard X--ray Survey
(Krivonos et al. 2012). All observations were made with the X--ray telescope on board 
\emph{Swift} and the data analysis was performed using the XRT pipeline included in the current 
version of HEASoft (v. 6.12).

\section{SWIFT J0958.0--4208}
This source is also reported in the \emph{Swift}/BAT 58--month catalogue\footnote{
available at http://heasarc.nasa.gov/docs/swift/results/bs58mon/.}.
XRT detects two X--ray objects within the IBIS/BAT error circles (see Figure~\ref{f1}, \emph{upper--left 
panel}):
\begin{itemize}
\item Source \#1: This is the brightest of the two detections, being detected
at 31$\sigma$ and 23$\sigma$ confidence level (c.l.) in the 0.3--10 keV energy range and above 3 keV,
respectively. The XRT position is R.A.(J2000) = 09$^{h}$57$^{m}$50.66$^{s}$ and
Dec.(J2000) = $-42^{d}08^{m}37.5^{s}$ (3$^{\prime\prime}$.6 uncertainty).
There are three USNO--B1.0 objects falling within the XRT positional uncertainty and one 2MASS source
(2MASS J09575064--4208355), with magnitudes $J$ $\sim$15.4, $H$ $\sim$14.9, and $K$ $\sim$14.9.

The X--ray data are well modelled using a black body component ($kT = 0.092^{+0.038}_{-0.041}$ keV) 
and a power law having a flat ($\Gamma = 0.77^{+0.10}_{-0.11}$) spectrum ($N_{\rm H(Gal)} =
1.65 \times 10^{21}$ cm$^{-2}$).
The observed 2--10 keV flux is $\sim$$4.7 \times 10^{-12}$ erg cm$^{-2}$ s$^{-1}$.

\item Source \#2: It is detected at 10.4$\sigma$ and 4.8$\sigma$ in the the 0.3--10 keV energy range 
and above 3 keV, respectively. It is located at R.A.(J2000) = 09$^{h}$57$^{m}$40.71$^{s}$ and
Dec.(J2000) = $-42^{d}08^{m}44.6^{s}$ (4$^{\prime\prime}$.0 uncertainty).
Within the XRT positional error, we find a USNO--B1.0 source (USNO--B1.0 0478--0250012) with
magnitudes $B$ $\sim$19.7 and $R$ $\sim$17.6--19.3.

The X--ray spectrum is fitted with a power law ($N_{\rm H(Gal)}  = 1.66 \times 10^{21}$ cm$^{-2}$)
having photon index $\Gamma= 1.51^{+0.26}_{-0.25}$ and an observed 2--10 keV flux of
$\sim$$2.8\times10^{-13}$ erg cm$^{-2}$ s$^{-1}$.
\end{itemize}

Although source \#1 is probably the more likely counterpart to SWIFT J0958.0--4208, we cannot exclude
at this stage contamination from the second object.

\section{SWIFT J1508.6--4953}
\begin{figure*}
\centering
\includegraphics[width=0.33\linewidth]{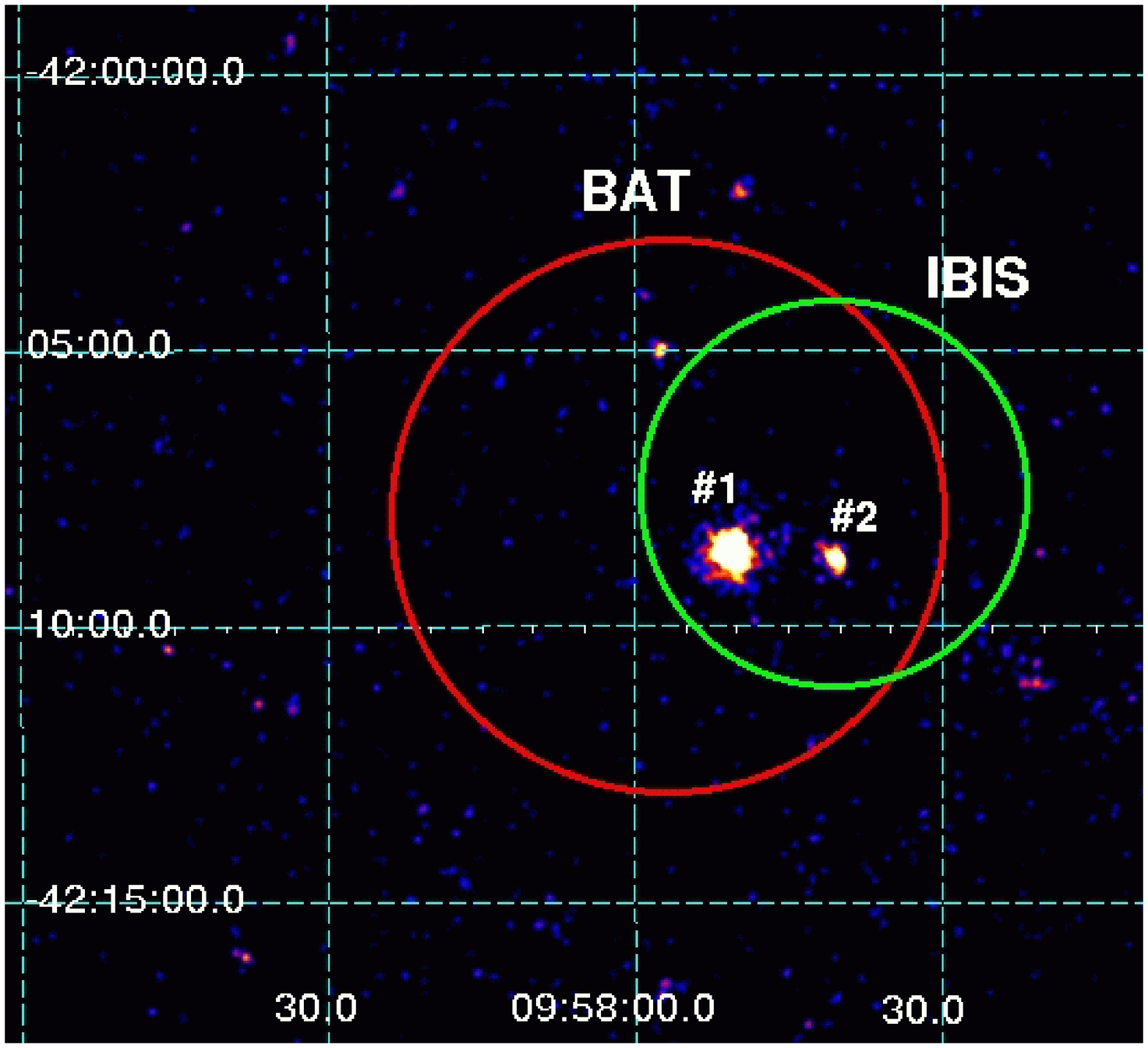}
\includegraphics[width=0.33\linewidth]{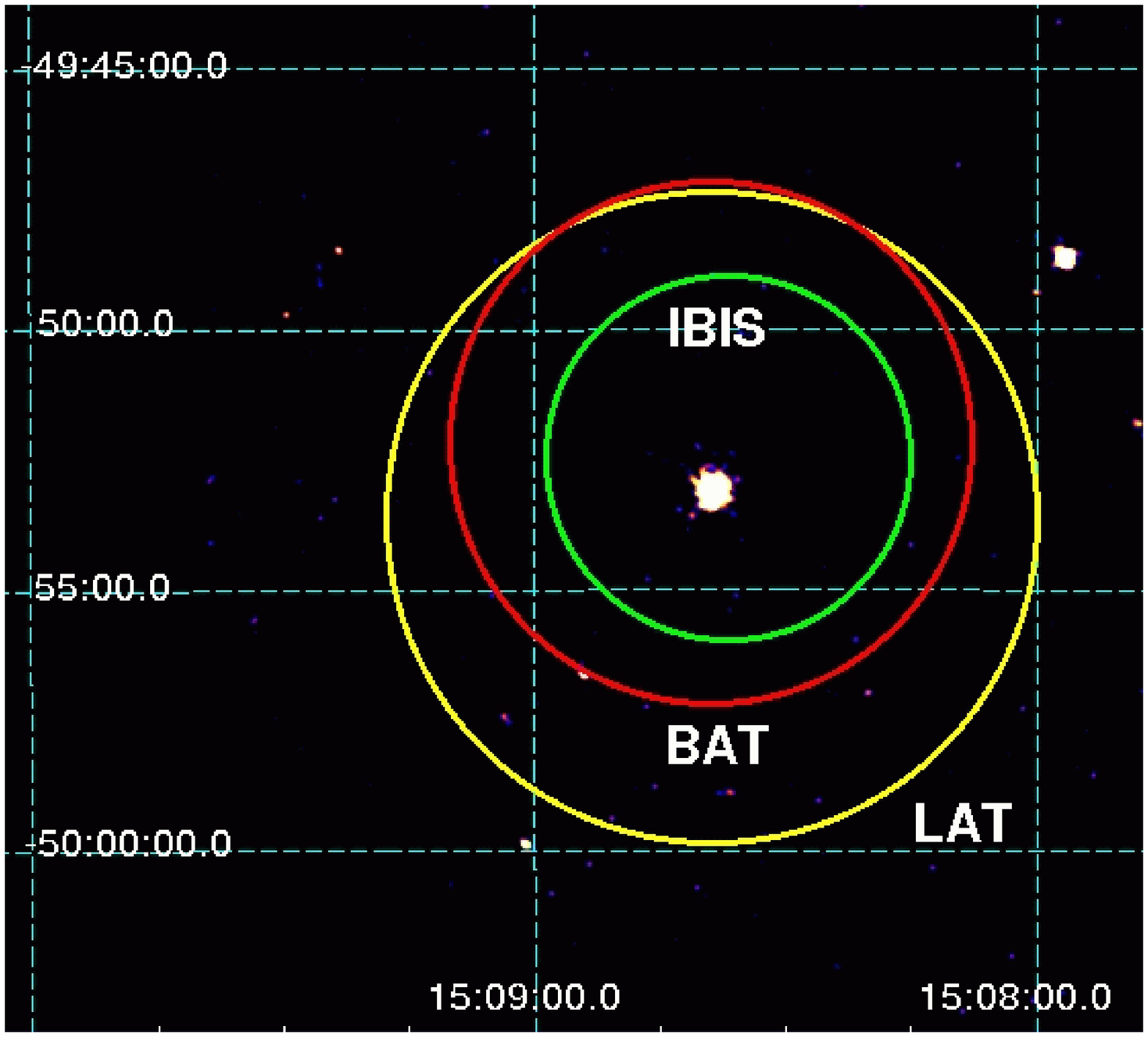}
\includegraphics[width=0.33\linewidth]{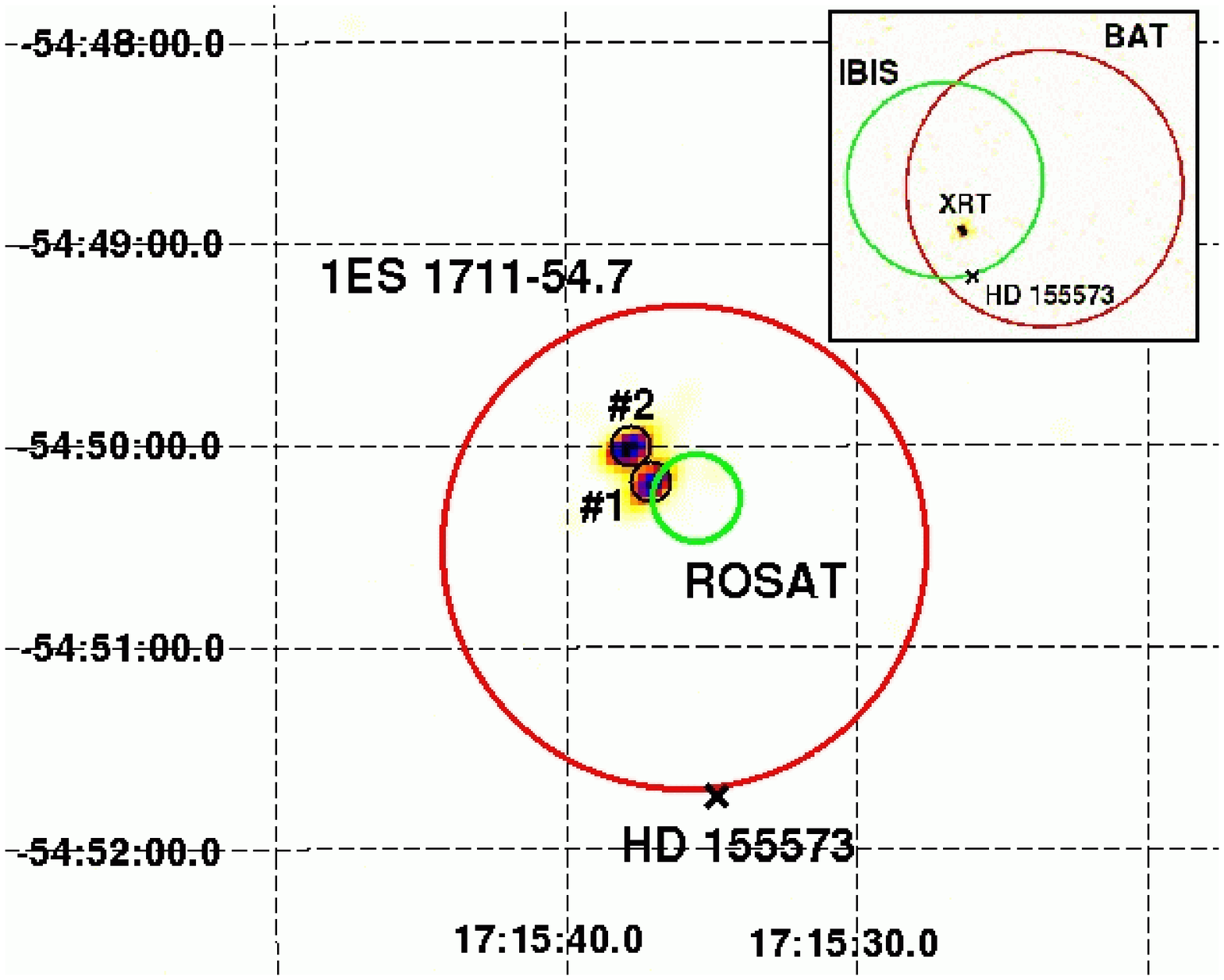}
\includegraphics[width=0.33\linewidth]{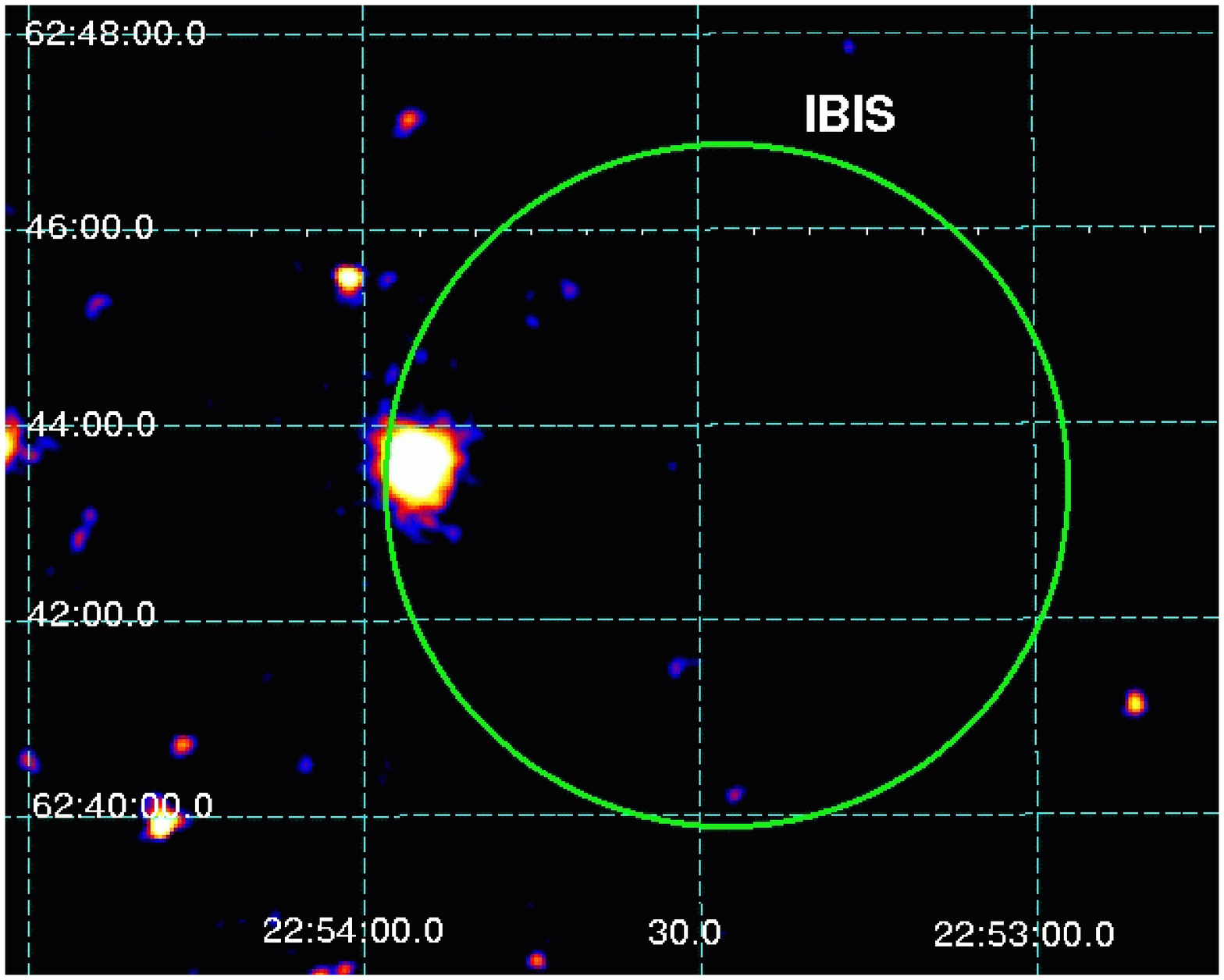}
\caption {0.3--10 keV XRT image of the region surrounding SWIFT J0958.0--4208 (\emph{upper--left 
panel}),
SWIFT J1508.6--4953 (\emph{upper--right panel}), IGR J17157--5449 (\emph{bottom--left panel}), and
IGR J22534+6243 (\emph{bottom--right panel}). See details in the text.}
\label{f1}
\end{figure*}

The IBIS source has been associated with the radio source PMN J1508--4953, reported as a GeV
emitter in the 2nd \emph{Fermi} catalogue (Nolan et al. 2012) and also listed in the
the 2nd \emph{Fermi} AGN catalogue (Ackermann et al. 2011) as an active galaxy of uncertain type.
There is only one X--ray source within 
the IBIS/BAT/LAT error circles (see Figure~\ref{f1}, \emph{upper--right panel}); this object is located at
R.A.(J2000) = 15$^{h}$08$^{m}$38.82$^{s}$ and
Dec.(J2000) = $-49^{d}53^{m}02.9^{s}$ (3$^{\prime\prime}$.6 uncertainty).
It is detected at 22.7$\sigma$ and 13$\sigma$ c.l. in the 0.3--10 keV
energy range and above 3 keV, respectively. Its position is compatible with that of the faint ROSAT
source 1RXS J150839.0--495304.

The XRT spectrum is well fitted with a power law having ($N_{\rm H(Gal)} = 2.04 \times 10^{21}$
cm$^{-2}$) a photon index $\Gamma = 1.41\pm 0.10$
and an observed 2--10 keV flux of $\sim$$2.8 \times 10^{-12}$ erg cm$^{-2}$ s$^{-1}$.

SWIFT J1508.6--4953 is most likely a Blazar since it is a radio loud object\footnote{$Log R = 
Log(L\_5GHz/L\_B)= 3.50$.} 
and has WISE colours ($W2-W3 = 3.15$ and $W1-W2 = 1.26$), typical of a gamma--ray emitting blazar
(Massaro et al. 2012) as mentioned above. \emph{Plank} data provide a spectrum
typical of a flat radio source (Tuerler at el. 2012), which combined with the fact that the LAT spectrum
has a photon index of 2.6, further suggests that SWIFT J1508.6--4953 could be a flat spectrum radio
quasar.

\section{IGR J17157--5449 also 2PBC J1715.2--5448}
As shown in the inset of the \emph{bottom--left panel} of Figure~\ref{f1}, there is only
one X--ray excess detected by XRT within the IBIS (green) and BAT (red) error circles. 
It is detected at 9.6$\sigma$ and 4.5$\sigma$ c.l. in the 0.3--10 keV energy range and above 3 keV,
respectively.
It is located at R.A.(J2000) =
17$^{h}$15$^{m}$37.39$^{s}$ and Dec.(J2000) = $-54^{d}50^{m}04.5^{s}$ (4$^{\prime\prime}$.3 uncertainty).

By zooming in on this region (larger image), we find that the X--ray source position is compatible 
with that of the bright ROSAT source 1RXS J171535.6--545015 (small green circle) and also falls within the 
positional uncertainty of an \emph{Einstein} Slew source 1ES 1711--54.7 (bigger red circle). In the 
light of these findings, the association between the star HD 155573 (black cross in the image) and the 
IBIS source, proposed by Krivonos et al. (2012), is ruled out. Indeed, although the star is located within 
the IBIS and BAT error circles, it is not detected by XRT and it lies far away from both the XRT and the 
\emph{Einstein} source detection.

A closer scrutiny of the XRT image indicates that the X-ray excess is composed of two distinct
objects (\#1 and \#2 in the image), both of which are still observed above 3 keV. By inspection of the
XRT image, we provide a tentative estimation of the position of each source (6$^{\prime\prime}$.0
uncertainty at 90\% c.l. assumed).
\begin{itemize}
\item Source \#1: R.A.(J2000) = 17$^{h}$15$^{m}$37.11$^{s}$ and Dec.(J2000) = $-54^{d}50^{m}10.7^{s}$.
Its position is compatible with two USNO--B1.0 objects: USNO--B1.0 0351--0716548 with $R$ $\sim$15.6 
and USNO--B1.0 0351--0716550 with $R$ $\sim$14.5--15.0.
\item Source \#2: R.A.(J2000) = 17$^{h}$15$^{m}$37.83$^{s}$ and Dec.(J2000) = $-54^{d}50^{m}00.2^{s}$.
It has no counterpart in publicly available databases.
\end{itemize}

Source \#1 and \#2 have comparable X-ray fluxes, which suggest that both objects could contribute to
the high energy emission seen by IBIS and BAT.

\section{IGR J22534+6243}

There is an X--ray source compatible with the 90\% IBIS error circle (see Figure~\ref{f1}, \emph{bottom 
right panel}), serendipitously detected during
the \emph{Swift}/XRT pointing of the gamma ray burst GRB060421. It has coordinates
R.A.(J2000) = 22$^{h}$ 53$^{m}$55.23$^{s}$ and Dec.(J2000) = $+62^{d}43^{m}38.0^{s}$
(3$^{\prime\prime}$.5 uncertainty), and is detected at 45$\sigma$ and 35$\sigma$ c.l. in the 0.3--10 keV 
energy range and above 3 keV, respectively. It is associated with the ROSAT
faint source 1RXS J225352.8+624354, which is still unidentified. Within the XRT positional
uncertainty we find a USNO--B1.0 object (USNO--B1.0 1527--0428738) having magnitudes $R$ $\sim$13 and
$B$ $\sim$15.5, which is also listed in the 2MASS catalogue (2MASS J22535512+6243368), with magnitudes
$J$ $\sim$11.6, $H$ $\sim$11.0, and $K$ $\sim$10.5.

The X-ray spectrum is well modelled with an absorbed ($N_{\rm H(Gal)} = 8.96 \times 10^{21}$
cm$^{-2}$, $N_{\rm H(intr)} = 1.05^{+0.25}_{-0.22} \times 10^{22}$ cm$^{-2}$)
power law ($\Gamma = 1.37^{+0.14}_{-0.13}$) having an observed 2--10 keV flux of  
$\sim$$3.4 \times 10^{-12}$ erg cm$^{-2}$ s$^{-1}$.

Follow--up observations/analyses further revealed the presence of an X--ray pulsation of 46.6s 
(Halpern 2012; Israel et al. 2012), which indicates a likely HMXB
nature for this object. This classification has been confirmed through optical follow--up 
observations (Masetti et al. 2012).

\section{Conclusions}
In this work, we show how follow--up observations in X--rays play a key role in searching
for counterparts of high energy emitters. The cross--correlation between the IBIS surveys and the
\emph{Swift}/XRT data archive allowed us to discuss the counterparts of four newly discovered
objects. We find that SWIFT J1508.6--4953 is most likely a Blazar, whereas IGR J22534+6243 is
probably a HMXB. As far as the other two objects are concerned, the lack of a unique X--ray counterpart 
makes the association less secure, and their nature can be inferred only by means of optical
follow--up observations of all likely counterparts.

\end{document}